\documentclass[a4paper,USenglish,cleveref, autoref]{lipics-v2019}

\usepackage{cite}
\usepackage{amsmath,amsfonts}
\usepackage{algorithmic}
\usepackage{xcolor}
\usepackage{booktabs}
\usepackage{pgfplots}
\pgfplotsset{compat=1.8}

\title{GPU Acceleration of Real-Time Control Loops}

\author{Mohamed A. Bamakhrama}{ASML, Netherlands}{}{}{}
\author{Alejandro Arrizabalaga}{ASML, Netherlands}{}{}{}
\author{Frank Overman}{ASML, Netherlands}{}{}{}
\author{Jean-Paul Smeets}{ASML, Netherlands}{}{}{}
\author{Kornel van der Sommen}{ASML, Netherlands}{}{}{}
\author{Remko van der Vossen}{ASML, Netherlands}{}{}{}
\author{John Wagensveld}{ASML, Netherlands}{}{}{}

\authorrunning{M. A. Bamakhrama et al.}

\Copyright{Mohamed A. Bamakhrama, Alejandro Arrizabalaga, Frank Overman, Jean-Paul Smeets, Kornel van der Sommen, Remko van der Vossen, and John Wagensveld}



\category{}

\relatedversion{}

\supplement{}

\acknowledgements{The authors would like to thank Peter Lindstrom and Matthew Larsen from Lawrence Livermore National Laboratory (LLNL) for their support with zfp.}

\nolinenumbers 

\hideLIPIcs  

\EventEditors{John Q. Open and Joan R. Access}
\EventNoEds{2}
\EventLongTitle{31st Euromico Conference on Real-Time Systems (ECRTS 2019)}
\EventShortTitle{ECRTS 2019}
\EventAcronym{ECRTS}
\EventYear{2019}
\EventDate{July 24--27, 2019}
\EventLocation{Stuttgart, Germany}
\EventLogo{}
\SeriesVolume{42}
\ArticleNo{23}

\begin{document}
\maketitle

\begin{abstract}
Extreme Ultraviolet (EUV) photolithography is seen as the key enabler for
increasing transistor density in the next decade.
In EUV lithography, 13.5 nm EUV light is illuminated through a reticle, holding a pattern to be printed, onto a silicon wafer.
This process is performed about 100 times per wafer, at a rate of over a hundred wafers an hour.
During this process, a certain percentage of the light energy is converted into
heat in the wafer.
In turn, this heat causes the wafer to deform which increases the overlay error,
and as a result, reduces the manufacturing yield.
To alleviate this, we propose a firm real-time control system that uses a wafer
heat feed-forward model to compensate for the wafer deformation.
The model calculates the expected wafer deformation, and then, compensates for
that by adjusting the light projection and/or the wafer movement.
However, the model computational demands are very high.
As a result, it needs to be executed on dedicated HW that can perform
computations quickly.
To this end, we deploy Graphics Processing Units (GPUs) to accelerate the
calculations.
In order to fit the computations within the required time budgets, we combine in a novel manner multiple techniques, such as compression and mixed-precision
arithmetic, with recent advancements in GPUs to build a GPU-based real-time control system.
A proof-of-concept implementation using NVIDIA P100 GPUs is able to deliver decompression throughput of 33 GB/s and a sustained 198 GFLOP/s per GPU for mixed-precision dense matrix-vector multiplication.

\end{abstract}

\section{Introduction}
Extreme Ultraviolet (EUV) photolithography is seen as the key enabler for
reducing the transistor dimensions, and hence, increasing transistor density in
the next decade.
EUV lithography has been in development for more than two decades. However,
EUV systems have been deployed in production only in the past five years.
As the adoption ramps up into high volume production, EUV lithography
manufacturers are focusing now on improving the technology further~\cite{vanSchoot:2017:SPIE}.
One of the improvements is increasing the power of the light dosage delivered to the wafer.
Naturally, a certain amount of the delivered power is converted into heat in the wafer.
As the power increases, the amount of power converted into heat increases too.
This extra heat poses a problem; it causes the wafer to deform which means that
the overlay error is increased.
As a result, this extra error leads to reduced manufacturing yield.
To alleviate this, a cooling hood is introduced to absorb the extra heat~\cite{vanSchoot:2017:SPIE}.
However, the cooling hood alone is not sufficient to reduce the overlay error to the target values.
As a solution, a \textbf{wafer heat feed forward (WHFF)} model is used in addition to the cooling hood~\cite{vandenHurk:2018:SICEISCS}.
The WHFF model is composed of two parts: (i) the \emph{thermal} model which computes the temperature of each point in the wafer after being exposed to a certain light dosage, and (ii) the \emph{deformation} model which computes the deformation in wafer shape due to the absorbed heat.
The WHFF model computes an accurate approximation of the temperature and deformation which are then sent to the actuators.
After that, the actuators perform corrections to the EUV light projection and/or wafer movement in order to minimize the overlay error.

One major issue in realizing the WHFF model is its huge computational requirements~\cite{vandenHurk:2018:SICEISCS}.
The WHFF model involves solving a heat dissipation equation (for the thermal model) together with a 2.5D mechanical deformation equation.
The heat dissipation equation is solved using Finite Elements Method (FEM) with a uniform mesh grid.
The temperature points resulting from solving the FEM formulation are applied to the 2.5D mechanical deformation equation to compute the deformation values in $x$, $y$, and $z$ axes.
To reach the required levels of accuracy, the thermal model uses a 2D mesh grid that has more than $360000$ points.
For a 300 mm wafer, this means computing a temperature point for every $0.25$ mm\textsuperscript{2}.
Once the thermal and deformation equations are formulated in matrix-vector form, the computational needs of the two models can be derived.
The performance needed to compute the WHFF model ``on-time'' to compensate for deformation is around $600$ GFLOP/s for all dimensions.
By ``on-time'', we mean computing the WHFF and delivering the resulting corrections to the actuators within a time window of 50 ms that allows those corrections to take effect.
If the corrections were to be sent after the 50 ms time window, then the mechanical parts would have moved and the corrections will have no impact.
That's why we classify the control system as \textbf{firm real-time}; missing a deadline is tolerable, however, it results in an increased overlay error (and hence lower yield) for the impacted dies and wafers.

In this paper, we present a realization of a firm real-time control system that runs the WHFF model on Commercial Off-The-Shelf (COTS) hardware components.
The proposed realization uses Graphics Processing Units (GPUs) to deliver the required throughput and memory bandwidth. 
In the past decade, there has been a lot of interest in using GPUs in embedded and industrial control systems~\cite{Elliott:2011:RTCSA,Hallmans:2012:ETFA,Hallmans:2013:ETFA,Chitchian:2013:TCST,Windmann:2016:INDIN,Maceina:2017:TNS,Maceina:2017:TNS2,Yang:2018:ECRTS}.
Most of this interest is driven by the proliferation of machine learning workloads that benefit significantly from acceleration on GPUs (e.g., \cite{Windmann:2016:INDIN}).
Acceleration of other compute-intensive control loops remains a largely unexplored arena despite the few attempts~\cite{Chitchian:2013:TCST,Maceina:2017:TNS,Maceina:2017:TNS2}.
In this work, we propose a novel solution for the acceleration of compute-intensive control models used in firm real-time systems.
The novelty of the proposed solution stems from the way in which it combines multiple domains (real-time control, high-performance computing, and digital signal processing) together with recent advancements in GPU technology in a unique way to solve a real industrial problem.
Several key aspects of our proposed solution can be summarized as follows:
\begin{itemize}
 \item \textbf{High-Performance + Real-Time}: The system has to deliver a throughput of 10 GFLOP every 50 ms per axis. Such a high throughput is characteristic of HPC systems used for scientific computing. At the same time, the 50 ms is a firm deadline; any violation of this deadline will render the computed data useless.
In our proposed system, we combine such high-performance with a tight firm real-time control loop.
 \item \textbf{High Bandwidth Memory}: The WHFF model needs to access around {400 GB/s per axis}.
 Traditionally, GPUs main drawback was the the limited memory bandwidth.
 However, this is starting to change with the advancement of High Bandwidth Memory (HBM,~\cite{JEDEC:2015:HBM}).
 HBMs use 3D stacked memories with Through-Silicon-Vias (TSVs) to form a memory die that is connected to the logic die via an interposer.
 The latest GPUs from NVIDIA (e.g., P100~\cite{NVIDIA:2018:P100} and V100~\cite{NVIDIA:2018:V100}) provide 16-32 GB of HBM with a bandwidth in the range of 700-900 GB/s.
 \item \textbf{Compression}: HBMs alone are not sufficient as the connection between system memory and HBM is still based on the rather slow PCI-Express interconnect.
 In order to overcome the PCI-Express bandwidth bottleneck, we compress all the data that goes over PCI-Express to GPU.
 With compression, we are able to achieve 4-10x reduction in the data volume transferred over PCI-Express.
 
 \item \textbf{Mixed-Precision Arithmetic \cite{Baboulin:2009:CPC}}: A key technique to maximize throughput on modern GPUs is to perform different operations using different precision depending on: (i) operation cost, (ii) accuracy needed, and (ii) rounding error introduced by each operation.
 In this work, we perform multiplications (costly, low rounding error) using single-precision, and reductions (cheap, high rounding error) using double-precision.
 Key advantages of this approach are: (i) the ability to store all the inputs as single-precision compressed arrays which reduces data transfer volume, and (ii) utilize both single- and double-precision execution units.

\end{itemize}

The rest of this paper is organized as follows. Section~\ref{sec:background} gives a detailed overview of the WHFF model and the environment in which it runs. Section~\ref{sec:related_work} gives an overview of the related work.
Section~\ref{sec:proposed_solution} describes the proposed system and SW implementation.
Section~\ref{sec:results} presents the results of evaluating a proof-of-concept implementation of the proposed system based on NVIDIA P100 GPUs. Finally, Section~\ref{sec:conclusion} finishes the paper with conclusions.

\section{Background} \label{sec:background}
In this section, we provide an overview of how the WHFF model works. 
For an in-depth mathematical description of the model, one can refer to \cite{vandenHurk:2018:SICEISCS}.

\subsection{Wafer Heat Feed-Forward Model}

\begin{figure}[t!]
 \centering
 \includegraphics[scale=1]{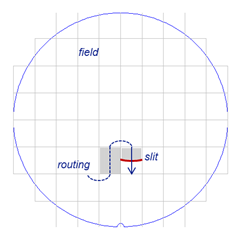}
 \caption{Wafer scan pattern}
 \label{fig:wafer}
\end{figure}

Figure~\ref{fig:wafer} illustrates the scan pattern of a wafer. 
A wafer is divided into \emph{fields}, where each field corresponds roughly to a single die.
Within a single field, the pattern is printed by illuminating EUV light through the reticle.
At a given time, the illuminated pattern corresponds to an arc-shaped \emph{slit}.
The scan speed can be either \emph{fast} or \emph{slow} depending on the \emph{energy dosage} to be delivered to the wafer. 
A \emph{fast} scan has shorter illumination period which means less energy and less time to scan a full field.
Conversely, a \emph{slow} scan has longer illumination period which means more energy and more time to scan a full field.
Furthermore, the scan time is classified into two inter-mixed stages: \emph{light} (i.e., when the illumninator is on), and \emph{dark} (i.e., when the illuminator is off).
{In a real EUV machine, the field fast scan is around 70 ms (light + dark) and the field slow scan is around 110 ms (light + dark).}
Unless mentioned otherwise, all real-valued data are stored as single-precision floating-point numbers.

The thermal model in the WHFF tracks the temperature development during the complete wafer scan.
As mentioned earlier, to reach the desired accuracy, the wafer is divided into uniform 2D mesh grid consisting of more than {$360000$} points. 
For every millisecond of the entire scan time (i.e., both light and dark stages), the thermal model has to be evaluated to compute the new temperature values.
In addition to the wafer, the thermal model takes also into account the wafer clamp (i.e., the part holding the wafer).
In total, there are around {$3.7 \times 10^5$} temperature values (stored in $\vec{T}_k$) that are updated in each iteration of the thermal model.
$\vec{T}_k$ is updated using Equation~\ref{eq:T_k}.
It is important to note that $\mathbf{A}$ is a sparse matrix, while $\mathbf{B}$ is a diagonal one.
\begin{equation} \label{eq:T_k}
\vec{T}_{k+1} = \mathbf{A} \cdot \vec{T}_k + \mathbf{B} \cdot \vec{u}_k
\end{equation}
After computing the temperature points, a thermal interpolation step is performed which reduces $\vec{T}_{k+1}$ to a smaller vector $\vec{S}_{k+1}$ that contains around {$2.5 \times 10^5$} values.

The 2.5D mechanical deformation model takes vector $\vec{S}_{k+1}$ (resulting from the thermal model) and multiplies it by matrix $\mathbf{C}_d$ at every light time step as shown in Equation~\ref{eq:delta_x}.
\begin{equation} \label{eq:delta_x}
 \forall k \in \text{light stage}: \quad \vec{\Delta}_d = \mathbf{C}_d \cdot \vec{S}_{k+1} \quad \forall d \in \{x, y, z\}
\end{equation}
The $\mathbf{C}_d$ matrix in Equation~\ref{eq:delta_x} is a dense matrix that describes, for every dimension $d \in \{x, y, z \}$, the deformation response of the complete mechanical setup, consisting of the wafer and the layers below it, to a variation in temperature at any point in the wafer surface.
The resulting $\vec{\Delta}_d$ contains (per dimension) the resulting deformations.
Equation~\ref{eq:delta_x} needs to be evaluated only for the light stage.
During the dark stage, the light is off and there is no extra heat transferred into the wafer.
However, keep in mind that the thermal model has to be evaluated for both light and dark stages to track the decrease of temperature during the dark stage.

Computing the real size of $\mathbf{C}_d$ (assuming single-precision floating-point) reveals that it is {229 GB per dimension $d$}.
This poses a huge challenge for implementing the WHFF on any HW platform since storing and processing such a huge matrix will be very costly.
To solve this challenge, we apply \emph{divide and conquer} approach.
Recall from Figure~\ref{fig:wafer} that the scanner moves sequentially across the fields and slits.
Therefore, one can reduce the size of $\mathbf{C}_d$ used in Equation~\ref{eq:delta_x} by computing \emph{only} the deformations impacting the field-under-scan.
This means that $\mathbf{C}_d$ can be reduced into $\mathbf{C}_{d,\text{field}}$.
This new smaller matrix has \emph{much fewer} rows than the original one (but the same number of columns as the original $\mathbf{C}_d$).
Furthermore, during the scan of a field, one can repeat the same strategy by computing only the deformations impacting the slit-under-scan using a matrix $\mathbf{C}_{d,\text{slit}}$ that is a subset of $\mathbf{C}_{d,\text{field}}$.
This means that, for every field, a single $\mathbf{C}_{d,\text{field}}$ is fetched, then multiple instances of $\mathbf{C}_{d,\text{slit}}$ are fetched from $\mathbf{C}_{d,\text{field}}$.
In other words:
$\mathbf{C}_{d,\text{slit}} \subset \mathbf{C}_{d,\text{field}} \subset \mathbf{C}_{d} $.
Computing the real sizes of these new smaller matrices (assuming single-precision) shows that $\mathbf{C}_{d,\text{field}}$ is {2067 MB} and $\mathbf{C}_{d,\text{slit}}$ is {361 MB}.

A complete pseudo-code summary of the WHFF model is depicted in Figure~\ref{fig:whff_algo}.
Based on this summary, we can compute the complexity of the WHFF model per scan millisecond.
One can easily observe that the WHFF model is dominated by \underline{GE}neralized \underline{M}atrix-\underline{V}ector (GEMV) multiplication which is a \underline{B}asic \underline{L}inear \underline{A}lgebra \underline{S}ubprograms (BLAS) Level-2 routine.
GEMV is an I/O bound algorithm with an arithmetic intensity \cite{Williams:2009:CACM} ranging from $\frac{1}{6}$ to $\frac{1}{2}$ asymptotically.
Having said that, we proceed now with computing the exact FLOP and I/O cost of the model.
Recall from linear algebra that dense matrix-vector product $\mathbf{A} \cdot \vec{b}$ of a matrix with $M$ rows and $N$ columns costs $MN$ multiplications and $M(N-1)$ summations. Hence, the total FLOP cost is $M(2N-1)$ FLOPs.
For sparse matrices, the FLOP cost is $M (2 \cdot \textsf{nnz}(\mathbf{A})-1)$ FLOPs, where $\textsf{nnz}(\mathbf{A})$ is the number of non-zero elements per row.
Let $t_l$ be the number of light time steps and $t_d$ be the number of dark time steps.
Then, the FLOP cost of the WHFF model per field is given by Equation~\ref{eq:flop_cost}.
\begin{equation} \label{eq:flop_cost}
\frac{\text{FLOPS}}{\text{Time Budget (ms)}} =  
  \underbrace{3}_{\text{per axis}}  \underbrace{ t_d \Big( M(2S - 1) \Big)}_{\text{Deformation Model}} + 
 \underbrace{ T (t_l + t_d) \Big( 2 \cdot \textsf{nnz}(\mathbf{A}) + 2 \cdot \textsf{nnz}(\mathbf{B}) -1 \Big)}_{\text{Thermal Model}} 
\end{equation}
where $T$, $M$, and $S$ are the dimensions specified in Figure~\ref{fig:whff_algo}.
The time budget in Equation~\ref{eq:flop_cost} is the time allocated for executing the model and it is set to around {50 ms} (for fast scans) and around {80 ms} (for slow scans).

If we substitute $T$, $M$, and $S$ in Equation~\ref{eq:flop_cost} with their real values, then we obtain the following lower bounds on the total WHFF computational requirements: {398.7 GFLOP/s} (for fast scans) and {587.2 GFLOP/s} (for slow scans).
However, one can easily note that the loop in line \ref{line:dim_loop} in Figure~\ref{fig:whff_algo} is fully parallelizable since there are no loop carried dependencies.
This allows us to speed-up the model execution by running all three iterations in parallel.
By doing that, we find that the throughput requirements \emph{per axis} are {150 GFLOP/s} (for fast scans) and {195 GFLOP/s} (for slow scans).

\begin{figure}[t!]
\hrule
\begin{algorithmic}[1]
 \REQUIRE Dark phase load
 \REQUIRE Wafer position
 \REQUIRE Heat loads impacting the wafer
 \REQUIRE Sparse matrix $\mathbf{A} \in \mathbb{R} ^{T \times T}$
 \REQUIRE Diagonal matrix $\mathbf{B} \in \mathbb{R} ^{T \times T}$
 \REQUIRE Dense matrices $\mathbf{C}_d \in \mathbb{R}^{K \times S}$, where $d \in \{x, y, z\}$ 
 \FOR{Every exposure scan}
  \STATE Fetch a sub-matrix $\mathbf{C}_{d,\text{field}} \in \mathbb{R}^{L \times S}$ corresponding to the field-under-scan from $\mathbf{C}_d$ \label{line:field}
  \FOR{Every millisecond during scan (light and dark)}
   \STATE Source term calculation (produces $\vec{u}_k \in \mathbb{R}^{T\times 1}$)
   \STATE Thermal Model (Equation~\ref{eq:T_k})
   \STATE Thermal Interpolation (produces $\vec{S}_k \in \mathbb{R}^{S \times 1} $)
   \IF{scan during light stage}
    \FOR{Every dimension $d \in \{x, y, z\}$} \label{line:dim_loop}
     \STATE Fetch a sub-matrix $\mathbf{C}_{d,\text{slit}} \in \mathbb{R}^{M \times S}$ from $\mathbf{C}_{d,\text{field}}$
     \STATE Deformation Model: $\vec{\Delta}_d = \mathbf{C}_{d,\text{slit}} \cdot \vec{S}_{k+1}$ \label{eq:delta_refined}
     \STATE Deformation Interpolation 
    \ENDFOR
   \ENDIF
  \ENDFOR
 \ENDFOR
 \RETURN Wafer deformation $\vec{\Delta}_d$ in $x$, $y$, and $z$
\end{algorithmic}
\hrule
\caption{A pseudo-code implementation of the WHFF model}
\label{fig:whff_algo}
\end{figure}

In a similar manner, one can compute the I/O cost of the WHFF model (I/O here means the number of accesses needed to fetch the data into the HW).
For dense matrix-vector multiplication, one needs to fetch $MN + N$ values and write $M$ values.
So in total, the I/O cost is $MN + N + M$ I/O operations (IOPS).
Therefore, the I/O cost of the WHFF model to scan a single field is given by:
\begin{equation} \label{eq:iops_slit_cost}
\frac{\text{IOPS}}{\text{Time (ms)}} = 
(t_l + t_d) T \Big( 2T + 3 \Big)  + 3 t_d \Big( MS + S + M \Big)
\end{equation}

\begin{figure}[t!]
 \centering
 \resizebox{!}{0.75\textwidth}{\includegraphics{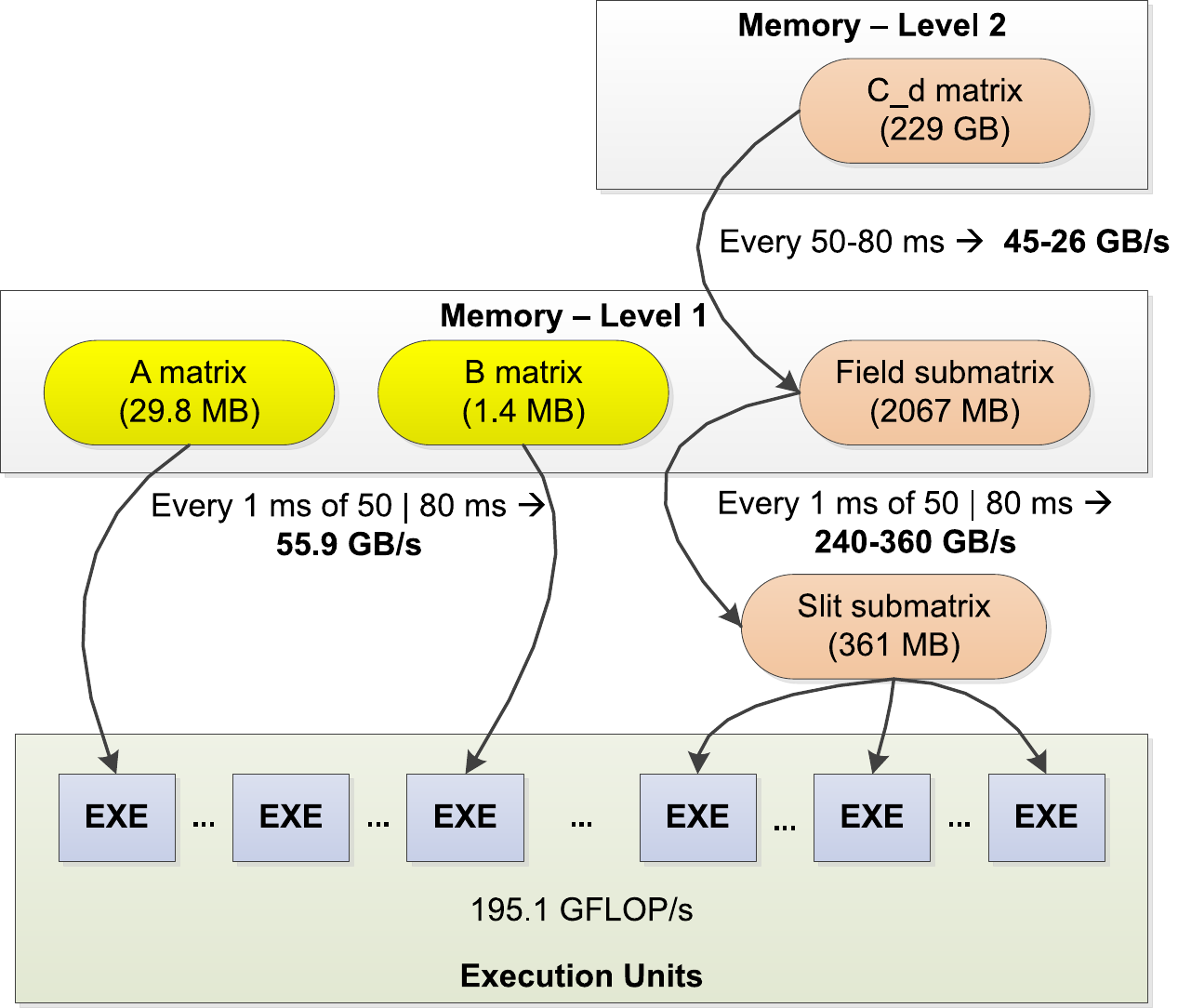}}
 \caption{Throughput and bandwidth requirements (per axis) for WHFF model}
 \label{fig:mapping}
\end{figure}

If we combine the actual throughput and memory bandwidth requirements of the WHFF model, then we obtain the picture shown in Figure~\ref{fig:mapping}.

\section{Related Work} \label{sec:related_work}
The past decade witnessed an increase in the interest of using General Purpose GPU (GPGPU) computing in embedded, real-time, and industrial systems~\cite{Elliott:2011:RTCSA,Hallmans:2012:ETFA,Chitchian:2013:TCST,Hallmans:2013:ETFA,Windmann:2016:INDIN,Maceina:2017:TNS,Maceina:2017:TNS2}.
In~\cite{Elliott:2011:RTCSA}, Elliot and Anderson discussed the applications that can benefit from GPUs and the constraints on using them.
In~\cite{Hallmans:2012:ETFA,Hallmans:2013:ETFA}, Hallmans et al.~discussed the potential of GPUs in embedded and industrial systems and the issues facing their adoption.

In~\cite{Chitchian:2013:TCST}, the first real attempt in accelerating a real-time control loop (involving a particles filter) on GPUs is proposed.
The proposed implementation was done on commercial GeForce cards and can be integrated in large industrial systems.
Similar to our proposed solution, \cite{Chitchian:2013:TCST} is targeting latencies below 50 ms.
However, a key difference between \cite{Chitchian:2013:TCST} and our solution is the model computational complexity and the amount of data needed to execute the model.
Particle filters are compute-bound and require small amounts of data compared to the dense matrices involved in solving the 3D mechanical deformation equation described in Section~\ref{sec:background}.

In~\cite{Windmann:2016:INDIN}, Windmann and Niggemann presented a method for fault detection in an industrial automation process.
The method incorporates a particle filter with switching neural networks in a fault detection method.
The execution time of the method was reduced from 80 s on CPUs to around 6 s on GPUs.
Compared to our proposed solution, \cite{Windmann:2016:INDIN} differs in: (i) their target latency is 100x our target latency, and (ii) the model is compute-bound with rather low memory bandwidth requirements.

In~\cite{Maceina:2017:TNS,Maceina:2017:TNS2}, Maceina and Manduchi presented an assessment of GPGPU in real-time control systems used in nuclear fusion reactors.
According to them, GPUs had limited success in real-time control due to: (i) \emph{lack of highly parallel real control applications}, and (ii) \emph{memory bandwidth bottleneck between system memory and GPU}.
They demonstrated that GPUs are very useful to accelerate multiple classes of compute-intensive control applications such as: (i) dense matrix-vector multiplication, (ii) image analysis, and (iii) synthetic magnetic measurements in magnetic confinement fusion.
Their first application (matrix-vector multiplication) is the same as the one that we accelerate in the WHFF.
However, key differences between our solution and the one in \cite{Maceina:2017:TNS} are:
\begin{itemize}
\item The size of the matrix and vector are much larger in our case.
To alleviate the memory bandwidth bottleneck, we use \emph{data compression}.
\item The use of mixed-precision arithmetic to increase the GPU utilization.
\item Another approach we use to alleviate the memory bandwidth bottleneck is the deployment of the latest GPUs equipped with High Bandwidth Memory (HBM,~\cite{JEDEC:2015:HBM}) such as NVIDIA's P100 and V100 GPUs.
\end{itemize}

\section{Proposed Solution} \label{sec:proposed_solution}
In this section, we outline the proposed solution to accelerate the WHFF model.
We start, in Section~\ref{sec:hw_selection}, by describing the criterion we followed to select GPUs from the different HW acceleration technologies available in the market.
Then, in Section~\ref{sec:system_arch}, we describe the overall system architecture.
After that, we describe the data compression scheme deployed in our system.
Next, in Section~\ref{sec:gemv}, we describe in details the mixed-precision matrix-vector multiplication scheme implemented on our system. 

\subsection{Why GPUs?} \label{sec:hw_selection}
\begin{figure}[t!]
 \centering
 \resizebox{!}{0.75\textwidth}{\sffamily \input{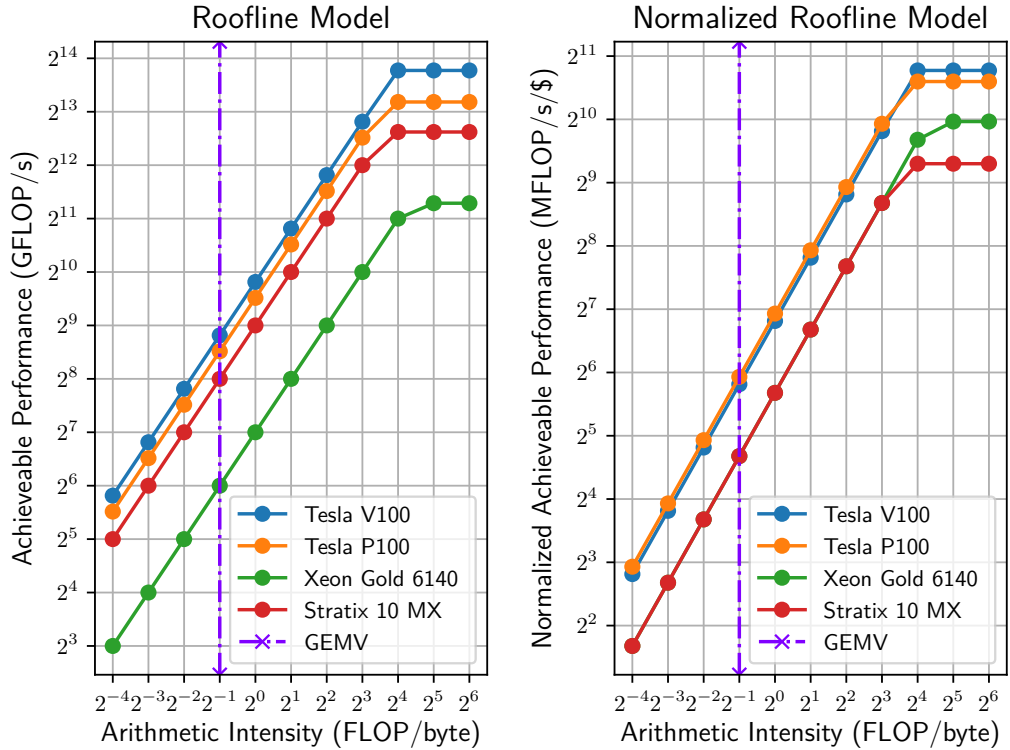}}
 \caption{Roofline model of modern computing HW platforms and the maximum achievable performance for single-precision GEMV on these HW platforms.}
 \label{fig:roofline}
\end{figure}

EUV lithography machines are unique industrial systems.
A single EUV lithography machine costs more than \$100~M, is power-rated at 1 MW, and has a lifetime span of 15-20 years.
A large portion of the machine power budget goes into the EUV light source, vacuum environment systems, and the mechanical parts.
Typically, when selecting electronics for EUV machines, \emph{cost efficiency} (i.e., throughput/\$) is often much more important than \emph{energy efficiency} (i.e., throughput/W).
Figure~\ref{fig:roofline} shows the Roofline model~\cite{Williams:2009:CACM} for a collection of high-end modern HW platforms covering CPUs, GPUs, and FPGAs together with a vertical line denoting the arithmetic intensity of GEMV.
The figure shows the \emph{canonical} Roofline (as defined in \cite{Williams:2009:CACM}) and the \emph{normalized} Roofline model (i.e., obtained through dividing throughput by price). The $y$-axis shows the maximum achievable single-precision performance. We observe the following:
\begin{enumerate}
\item HW with HBM2 memory (i.e., Teslas and Stratix) offers the best performance in the canonical Roofline.
\item FPGAs become worse in MFLOP/s/\$ compared to GPUs and CPUs. This has to do with the high price tag of HBM2 enabled FPGAs.
\item We observe an "inversion" between P100 and V100 in the normalized Roofline (i.e., P100 is better than V100). V100 is a newer card with higher bandwidth. However, the increase in bandwidth (and hence throughput) is not worth it if you normalize throughput by cost for IO bound applications.
\end{enumerate}

\subsection{System Architecture} \label{sec:system_arch}
\begin{figure}[t!]
 \centering
 \includegraphics[scale=0.7]{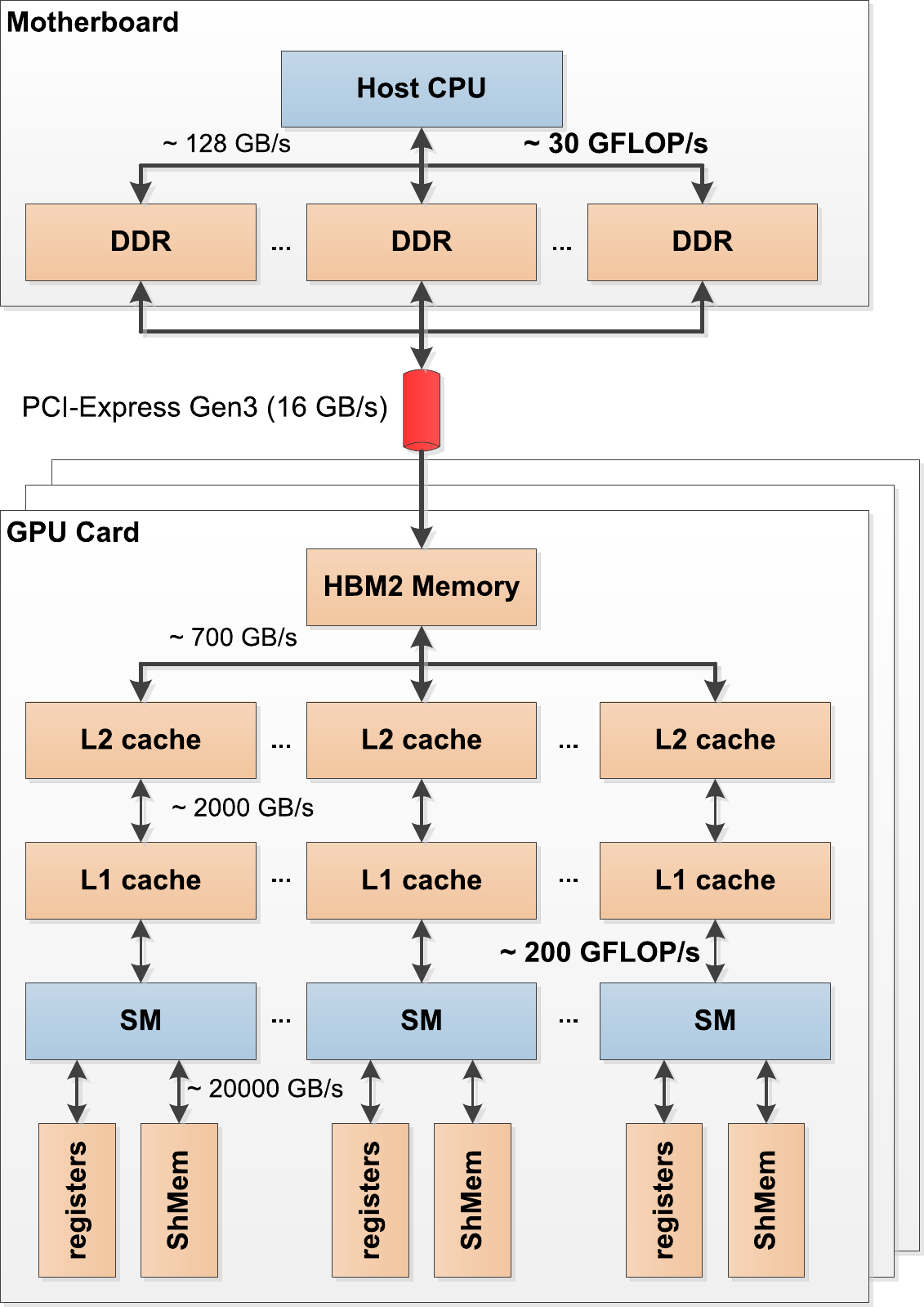}
 \caption{Overall System Architecture. Each GPU has its own dedicated 16-lane PCI Express Gen3 connection to the system memory. The host CPU is assumed to have enough PCI lanes (e.g., 48) to have a direct point-to-point connection with each GPU (i.e., no PCI-Express switch)}
 \label{fig:system_arch}
\end{figure}

If we take the requirements shown in Figure~\ref{fig:mapping} and map them to actual COTS HW, then we obtain the system architecture shown in Figure~\ref{fig:system_arch}.
The mapping in Figure~\ref{fig:mapping} shows the need for "level 1 memory" that can store $>$ {2067 MB} and support more than {410 GB/s}.
At the time of writing, the only COTS memories capable of realizing Level-1 memory are HBM and Graphics DDR (GDDR).
If we combine the FLOP requirements with Level-1 memory requirements, then we find that NVIDIA's V100 and P100 cards satisfy both requirements.
For level-2 memory, the $\mathbf{C}_d$ matrix is so huge ($>$ {200 GB}) that it can be stored only in the system DDR memory or permanent storage.
However, the interconnection from the system memory/storage to the GPU is slower than the required bandwidth as it is based on PCI-Express\footnote{We exclude NVIDIA's NVLink because it is a proprietary interconnect with no clear end-of-life strategy and is supported only on IBM POWER CPUs.} Gen3 (16 GB/s vs. 45 GB/s).
To tackle this problem, we deploy compression on the data transferred over PCI Express.
Compression is able to achieve compression ratios in the range of 4-10x.
A factor 4 allows the bandwidth requirement to drop from 45 GB/s to 11.25 GB/s which is well-below PCI-Express maximum bandwidth.
In the final system, three GPU cards are used where each card processes one dimension of the 3D mechanical deformation model (i.e., one iteration of the loop in line~\ref{line:dim_loop} in Figure~\ref{fig:whff_algo}).
Each card is connected to the system memory using a 16-lane PCI-Express Gen3 interconnect.

\subsection{Data Compression}
\begin{figure}[t!]
 \centering
 \resizebox{!}{0.35\textwidth}{\sffamily \includegraphics{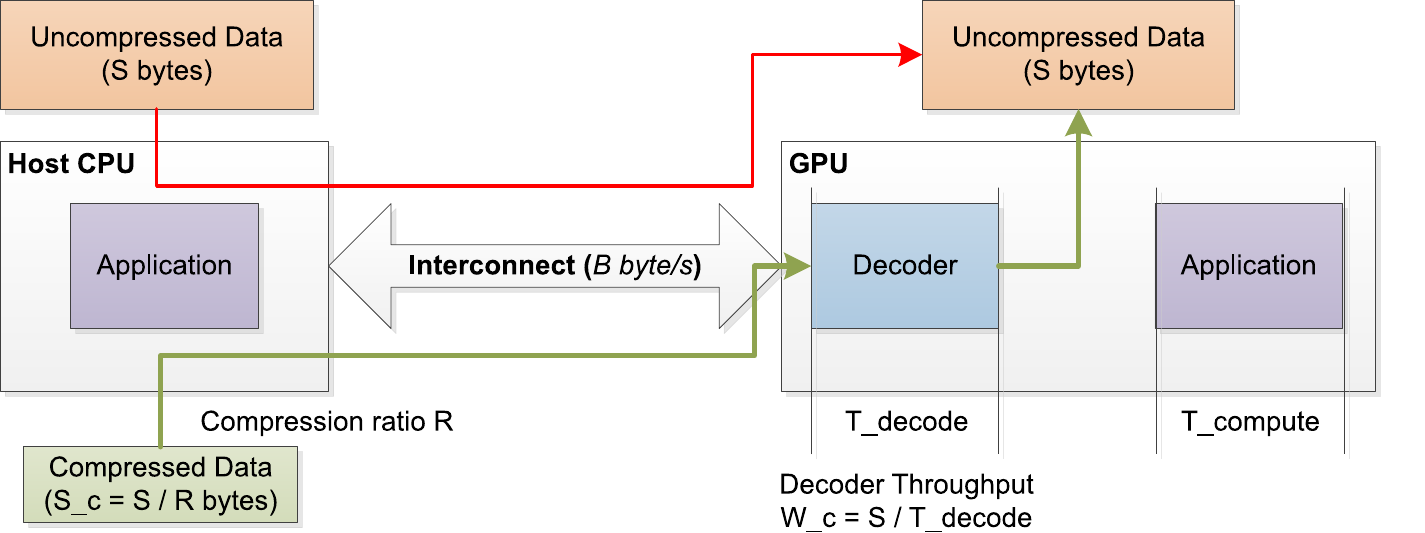}}
 \caption{The system model with data compression}
 \label{fig:compression_cost}
\end{figure}

As mentioned in Section~\ref{sec:system_arch}, we apply compression on the data transferred over PCI Express bus to the GPUs. 
Figure~\ref{fig:compression_cost} illustrates how compression is deployed.
The interconnect has a fixed bandwidth (denoted by $B$ and measured in byte/s). Suppose that we have a block of data consisting of $S$ bytes that needs to be transferred from the host CPU to the GPU.
We assume that data transfers are pipelined with computations \cite{NVIDIA:2018:Best}.
Without compression, it would take $T_\text{transfer} = S/B$ seconds to transfer the data from the host to GPU (red line in Figure~\ref{fig:compression_cost}).
With compression, the data is reduced in size by a compression factor $R$ which results in a compressed piece of data with size $S_c = \frac{S}{R}$ bytes. This means that the transfer time is reduced by a factor $R$ (green line in Figure~\ref{fig:compression_cost}).
Let $T_\text{compute}$ be the time needed to for the application computations and $T_\text{decode}$ be the time needed to decode the compressed data.
Then, Table~\ref{tab:comp_metrics} shows the pipeline period and latency of the total system with and without compression.
Note that in bandwidth-bound applications (e.g., GEMV), $T_\text{transfer}$ is much larger than $T_\text{compute}$ (i.e., $T_\text{transfer} \gg  T_\text{compute}$).
\begin{table}[t!]
 \centering
 \caption{Pipeline period and latency of the system with and without applying data compression}
 \begin{tabular}{|c|c|c|}
  \toprule
   & \textbf{Without compression} & \textbf{With compression} \\
  \midrule
  Pipeline Period (s) & $ \max \big( T_\text{transfer}, T_\text{compute} \big) $  & $\max \big( \frac{T_\text{transfer}}{R}, T_\text{compute} + T_\text{decode} \big) $ \\
  Latency (s) & $ T_\text{compute} + T_\text{transfer}$ & $ T_\text{compute} + \frac{T_\text{transfer}}{R} + T_\text{decode} $ \\
  \bottomrule
 \end{tabular}
 \label{tab:comp_metrics}
\end{table}
For compression to be beneficial, the following must hold:
\begin{enumerate}
 \item The decompression throughput (denoted by $W_c$ and measured in byte/s) must be larger than or equal to the interconnect bandwidth. $W_c$ is equal to $S / T_\text{decode}$. We show later in Section~\ref{sec:results} that we are able to get $W_c$ on NVIDIA P100 that is in the range of 30-35 GB/s which is higher than PCI-Express Gen3 16 GB/s maximum bandwidth.
 \item The GPU must have enough resources to accommodate both the application and the decompressor. This is true in our case as we do not run any other applications beside the WHFF on the GPU.
 \item The total latency after introducing decompression is still within the latency budget of the application. In other words, $T_\text{decode} + T_\text{compute} + T_\text{transfer}/R < T_\text{budget}$. This is also true in our application. Moreover, in our application, compression can be done offline (i.e., before application starts) which means that the encoding part is completely out of the time budget.
\end{enumerate}

We choose zfp~\cite{Lindstrom:2014:TVCG} as a compression scheme.
zfp is a compression algorithm that operates on integer and floating-point data. It can process data sets that are organized as 1D, 2D, 3D and 4D. 
zfp provides three modes of operation:
\begin{enumerate}
 \item \textbf{Fixed accuracy}: In this mode, the compressor accepts the uncompressed data and a \emph{tolerance}. Then, it produces a compressed data stream which, compared to the original data, has an absolute difference that is less than or equal to \emph{tolerance}. If the tolerance is set to zero, then zfp will operate in \emph{near lossless mode}.
 \item \textbf{Fixed precision}: In this mode, the compressor is controlled by specifying the number of most significant bits that the compressor must preserve from the original data into the compressed data.
  \item \textbf{Fixed rate}: In this mode, the compressor is controlled by specifying a fixed \emph{Bit-Per-Value (BPV)} rate. Then, the encoder encodes each floating point value using \emph{BPV} bits.
\end{enumerate}

\begin{table}[t!]
 \centering
 \caption{Results of compressing $\mathbf{C}_d$ under different modes of zfp. RMSE stands for Root Mean Square Error. NRMSE is Normalized RMSE. PSNR is Peak Signal-to-Noise Ratio.}
\resizebox{\textwidth}{!}{
 \begin{tabular}{|c|c|c|c|c|c|c|c|}
  \toprule
  \textbf{Mode} & \textbf{Axis} & \textbf{Bit/value} & \textbf{Rate} & \textbf{RMSE} & \textbf{NRMSE} & \textbf{Max. pointwise error} & \textbf{PSNR} \\
  \midrule
  \multirow{3}{*}{Fixed rate (8)} & $x$ & 8 & 4 & $3.510\times 10^{-13}$ & $2.36 \times 10^{-5}$ & $1.522 \times 10^{-10}$ & 86.54 \\
  & $y$ & 8 & 4 & $2.113 \times 10^{-13}$ & $1.95 \times 10^{-5}$ & $1.122 \times 10^{-10}$ & 88.19 \\
  & $z$ & 8 & 4 & $6.204 \times 10^{-13}$ & $2.77 \times 10^{-5}$ & $2.900 \times 10^{-10}$ & 85.14 \\ \midrule
  \multirow{3}{*}{Fixed-accuracy ($< 10^{-12}$)} & $x$ & 2.635 & 12.14 & $4.326\times 10^{-14}$ & $2.90 \times 10^{-6}$ & $3.974 \times 10^{-13}$ & 104.72 \\
  & $y$ & 2.579 & 12.41 & $4.526 \times 10^{-14}$ & $4.17 \times 10^{-6}$ & $4.081 \times 10^{-13}$ & 101.58 \\
  & $z$ & 3.38 & 9.47 & $4.942 \times 10^{-14}$ & $2.21 \times 10^{-6}$ & $4.107 \times 10^{-13}$ & 107.11 \\ \midrule
  \multirow{3}{*}{Fixed-precision (17 bits)} & $x$ & 7.6 & 4.21 & $3.925\times 10^{-15}$ & $2.63 \times 10^{-7}$ & $1.156 \times 10^{-12}$ & 125.57 \\
  & $y$ & 8.073 & 3.96 & $4.090 \times 10^{-15}$ & $3.77 \times 10^{-7}$ & $1.273 \times 10^{-12}$ & 122.46 \\
  & $z$ & 7.575 & 4.22 & $2.142 \times 10^{-14}$ & $9.55 \times 10^{-7}$ & $4.791 \times 10^{-12}$ & 114.38 \\
  \bottomrule
 \end{tabular}}
 \label{tab:zfp_comp_error}
\end{table}

Table~\ref{tab:zfp_comp_error} highlights the results of compressing $\mathbf{C}_d$ (from a mesh of $150 \times 200$) under the different modes of zfp.
It shows that fixed accuracy gives the best compression ratio (\textasciitilde 10x) while fixed-precision achieves the best NRMSE and PSNR.
Fixed-rate is very deterministic in timing and bandwidth, at the cost of much larger error.

The next step is to investigate the error introduced by the \emph{changed} decompressed data on the deformations computed by the WHFF model.
Upon evaluating the error introduced by compression on the final Quantity of Interest (QoI), we found that the error in deformations is $0.1\%$ of the deformations computed by the original $\mathbf{C}_d$ matrix (in fixed accuracy and precision modes).
Such a low error in the final QoI is acceptable as long as it is below 1\%.

\subsection{Mixed-Precision Matrix-Vector Multiplication} \label{sec:gemv}
As mentioned earlier, the core of the WHFF model is dense matrix-vector multiplication (i.e., GEMV).
Therefore, it is very critical to accelerate GEMV as much as possible.
Most HW vendors provide an optimized version of BLAS routines for their platforms.
For example, Intel provides Intel's Math Kernel Library (MKL, \cite{Intel:2018:MKL}) for the x86 platform, while NVIDIA provides cuBLAS library \cite{NVIDIA:2018:cuBLAS} for its GPUs.
In the WHFF, $\mathbf{C}_{d,\text{slit}}$ is a very wide matrix with {378 rows and 256000 columns}.
Such a wide matrix poses an interesting challenge.
In $\mathbf{A} \cdot \vec{b} = \vec{r}$, each element of $\vec{r}$ is computed using the dot operator as:
\begin{equation} \label{eq:dot}
r_i = \sum_{j=1}^{N} A(i,j) b_j
\end{equation}
If $\mathbf{A}$ is very wide, then the summation in Equation~\ref{eq:dot} becomes a very long reduction sequence.
If single-precision floating-point numbers are used, then for an $\mathbf{A}$ with a width of {256000} elements (the same as $\mathbf{C}_{d,\text{slit}}$), we lose 17 bits of the mantissa due to rounding errors \cite{Goldberg:1991:CSR}.
This leaves us with only 6 "un-contaminated" bits.
Therefore, it is essential to use double-precision for performing the reduction.
On the other hand, performing the multiplication in Equation~\ref{eq:dot} using single-precision is very safe (in terms of rounding error) and much faster than double-precision.
Therefore, we use a mixed-precision implementation of GEMV, where: (i) the inputs and outputs are both single-precision arrays, (ii) multiplication is performed using single-precision, and (iii) reduction is performed by casting the product from (ii) and adding it to a double-precision \texttt{sum} variable.
A sequential mixed-precision GEMV in C is shown in Figure~\ref{fig:naive_mp_gemv}.
\lstset{basicstyle=\footnotesize\ttfamily}
\begin{figure}[t!]
\begin{lstlisting}[frame=single,language=C,tabsize=2]
/* float is single-precision (32 bits)
   and double is double-precision (64 bits) */
void gemv(const float * __restrict__ M,
          const float * __restrict__ V,
                float * __restrict__ R)
{
 int x, y;
 double sum;
 float product;

 for (y = 0; y < HEIGHT; y++) {
  sum = 0.0;
  for (x = 0; x < WIDTH; x++) {
   product = (float)M[y*WIDTH + x] * (float)V[x];
   sum += (double) product;
  }
  R[y] = (float) sum;
 }
}
\end{lstlisting}
\caption{Mixed-precision sequential implementation of GEMV in C}
\label{fig:naive_mp_gemv}
\end{figure}

In this work, we implemented two variants of GEMV:
\begin{enumerate}
 \item \textbf{CPU-MP}: Implemented manually by auto-vectorizing a sequential version of mixed-precision GEMV using an auto-vectorizing compiler and targeting AVX-2 and OpenMP.
 \item \textbf{GPU-MP}: A manual mixed-precision implementation of GEMV (using CUDA and OpenCL).
\end{enumerate}

In the following subsections, we explain each variant in details.
\subsubsection{Variant 1: CPU-MP}
The CPU variant is based on a the sequential implementation shown in Figure~\ref{fig:naive_mp_gemv}.
We applied one modification to the code shown in Figure~\ref{fig:naive_mp_gemv} to enable auto-vectorization.
The change deals with telling the compiler that the input pointers are allocated using aligned memory allocators (e.g., POSIX's \texttt{posix\_memalign}). 
Aligning the pointers in memory enables the compiler to generate efficient vector code.
In addition, we also parallelized the outer loop over different cores using OpenMP.
The core of the GEMV kernel is vectorized nicely using GCC into AVX instructions as shown in Figure~\ref{fig:gemv_avx}.
\lstset{basicstyle=\footnotesize\ttfamily}
\begin{figure}[t!]
\begin{lstlisting}[frame=single,language=C,tabsize=2]
vmovaps 			(%rdx,%rax,1),%ymm0
vmulps 				(%rcx,%rax,1),%ymm0,%ymm0
add    				$0x20,%rax
vextractf128	$0x1,%ymm0,%xmm1
vcvtps2pd 		%xmm0,%ymm0
vcvtps2pd 		%xmm1,%ymm1
vaddpd 				%ymm0,%ymm1,%ymm0
vaddpd 				%ymm0,%ymm2,%ymm2
cmp    				$0x8000,%rax
\end{lstlisting}
\caption{AVX instructions implementing the product and summation inside GEMV}
\label{fig:gemv_avx}
\end{figure}
We can easily see that the compiler produces single-precision instructions (\texttt{vmulps}) for the multiplication and double-precision ones (\texttt{vaddpd}) for the addition.
The conversion from single to double precision is done using the \texttt{vcvtps2pd} instruction.

\subsubsection{Variant 2: GPU-MP}
The second variant is implemented using CUDA as shown in Figure~\ref{fig:cuda_gemv}.
It applies, next to mixed-precision arithmetic, \emph{memory coalescing} and \emph{parallel reduction} \cite{NVIDIA:2018:Best}.
Both techniques are widely used in GPUs to achieve higher performance.
In addition, it uses the GPU shared memory to achieve very short access time to the \texttt{partial\_sums} array.

\lstset{basicstyle=\footnotesize\ttfamily}
\begin{figure}[t!]
\begin{lstlisting}[frame=single,language=C,tabsize=2]
/* gemv multiplies matrix M (H x W) by
   vector V (W x 1) and stores the result
   in vector R (H x 1). */
__global__ void gemv(const real_t * __restrict__ M,
                     const real_t * __restrict__ V,
                           real_t * __restrict__ R)
{
 int x;
 __shared__ double partial_sums[THREAD];
 double sum = 0.0;
 float product;

#pragma unroll
 for(x = threadIdx.x; x < W; x += blockDim.x) {
  product = (float)M[blockIdx.x*W+x] * (float)V[x];
  sum += (double) product;
 }
 partial_sums[threadIdx.x] = sum;
 __syncthreads();

 // Perform parallel reduction on partial_sums.
 // Code is omitted for brevity. The result
 // of the parallel reduction is stored in
 // partial_sums[0]

 if(threadIdx.x == 0)
  R[blockIdx.x] = partial_sums[0];
}
\end{lstlisting}
\caption{Mixed-precision CUDA implementation of GEMV}
\label{fig:cuda_gemv}
\end{figure}

\section{Evaluation Results} \label{sec:results}
We have implemented a Proof-of-Concept (PoC) version of the WHFF model that runs the model for a single dimension.
The PoC HW uses Intel Xeon E5-2660 V4 clocked at 2.0 GHz interconnected to an NVIDIA P100 (16 GB HBM) using 16-lane PCI-Express Gen3.
For the SW, we use RedHat Enterprise Linux (RHEL) 7 together with Intel MKL 11.3.1.150, GCC 4.9.3, and CUDA 8.0.61.

In the PoC, we evaluated the two mixed-precision matrix-vector multiplication schemes on our proposed system.
To judge the performance of our two variants implemented in Section~\ref{sec:gemv}, we use the following two standard BLAS implementations of gemv:
\begin{enumerate}
 \item \textbf{Intel MKL}: Uses GEMV from Intel's Math Kernel Library (MKL) running on the host CPU.
 \item \textbf{cuBLAS}: Uses GEMV from NVIDIA's cuBLAS library running on the GPU.
\end{enumerate}
It is important to note that the two BLAS implementations do \emph{not} support mixed-precision arithmetic for BLAS Level-2 routines. 
Therefore, we use the double-precision version of GEMV (called DGEMV in contrast to the single-precision version SGEMV).

\subsection{GEMV Performance}
\begin{figure}
 \centering
 \footnotesize
 \resizebox{\columnwidth}{!}{
 \begin{tikzpicture}
\begin{axis}[
    ybar,
    width=\columnwidth,
    ymajorgrids,
    enlargelimits=0.15,
    legend style={at={(0.05,0.95)}, anchor=north west},
    ylabel={GFLOP/s},
    symbolic x coords={MKL, cuBLAS, CPU-MP, GPU-OpenCL, GPU-CUDA},
    xticklabels={MKL (DGEMV), cuBLAS (DGEMV), CPU-MP (AVX+OpenMP), GPU-MP (OpenCL), GPU-MP (CUDA)},
    xtick=data,
    xticklabel style={rotate=45,anchor=east},
    ]
\addplot coordinates {(MKL,11.69) (cuBLAS,14.59) (CPU-MP,23.14) (GPU-OpenCL, 164.33) (GPU-CUDA, 198.3)};
\addplot coordinates {(MKL,14.29) (cuBLAS,99.69) (CPU-MP,17.64) (GPU-OpenCL,16.79) (GPU-CUDA,65.9)};
\addplot coordinates {(MKL,14.23) (cuBLAS,87.83) (CPU-MP,21.61) (GPU-OpenCL,216.1) (GPU-CUDA,202.62)};
\legend{Wide Matrix (378$\times$256K), Tall Matrix (256K$\times$378),Square Matrix (10K$\times$10K)}
\end{axis}
\end{tikzpicture}}
 \caption{Results of evaluating different GEMV variants under different matrix layouts}
 \label{fig:results_throughput}
\end{figure}
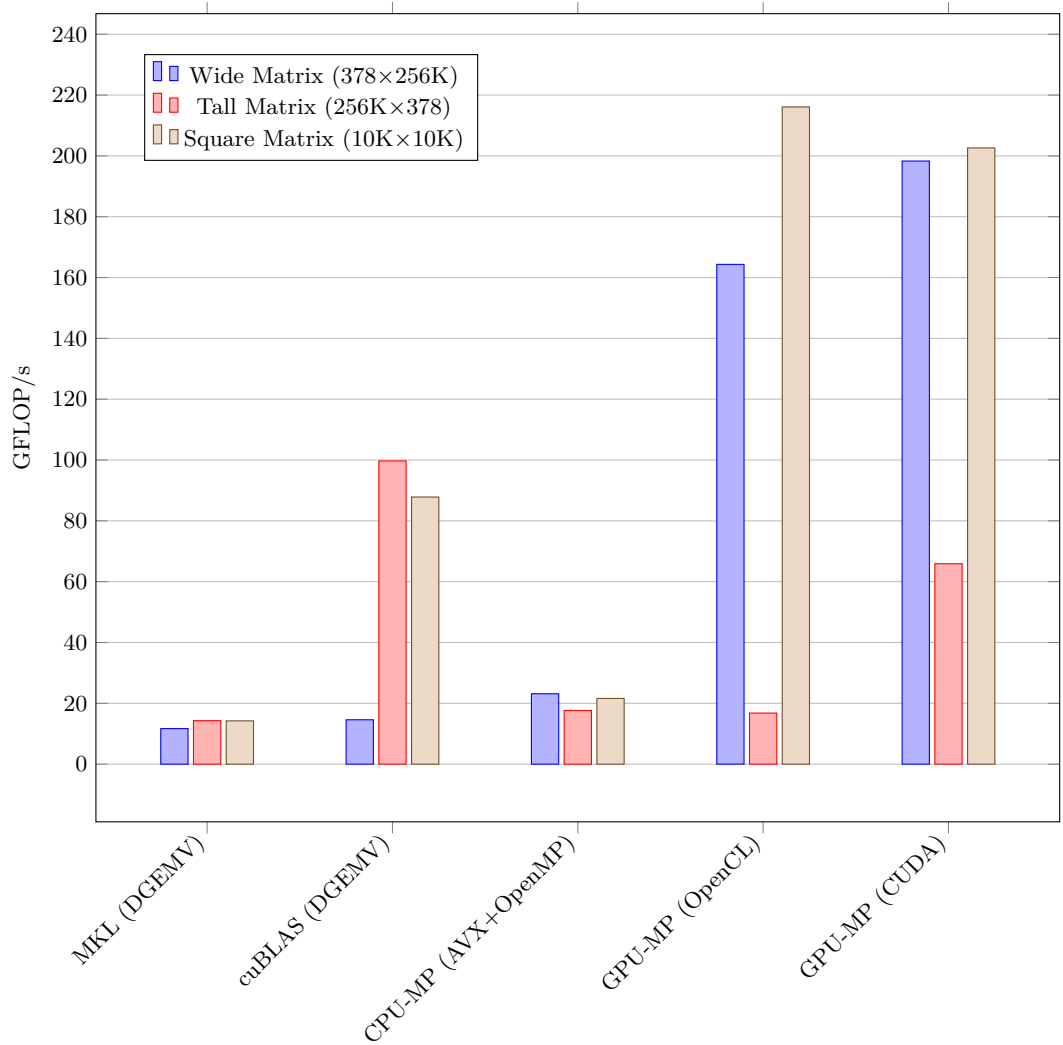

Figure~\ref{fig:results_throughput} shows the throughput (in GFLOP/s) of the different GEMV variants.
The first observation is that all vendor-provided variants fall short of the 195 GFLOP/s threshold shown in Figure~\ref{fig:mapping} which is needed to execute the WHFF model within its time budget.
The maximum achieved performance on P100 was obtained using the GPU-MP (CUDA) variant which is able to provide a sustained {198 GFLOP/s}.
This finding is interesting since the common wisdom is to use vendor-provided routines (e.g., MKL and cuBLAS) as they represent the most optimized implementation for the underlying HW.
However, in this specific case, using a custom mixed-precision implementation tailored for the application proved to be the right choice.

Another key observation from Figure~\ref{fig:results_throughput} is the better performance of CUDA over OpenCL for wide and tall matrices.
For square ones, OpenCL is slightly better.
Another interesting observation is the poor performance of cuBLAS on wide matrices.
This poor performance is due to the fact that CUDA GEMV is parallelized for the row direction with 1D thread-block and the row dimension is too small, in wide matrices, compared to the number of CUDA cores and SMs.

\subsection{Decompression Performance}
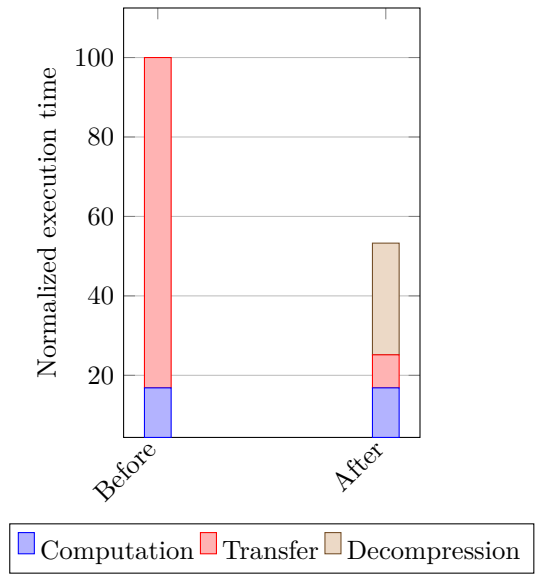
\begin{figure}[t!] 
\centering
\begin{tikzpicture}
\begin{axis}[
    ybar stacked,
    x=3cm,
    ymajorgrids,
    enlargelimits=0.15,
    legend style={at={(0.5,-0.20)}, anchor=north,legend columns=-1},
    ylabel={Normalized execution time},
    symbolic x coords={Before, After},
    xtick=data,
    x tick label style={rotate=45,anchor=east},
    ]
\addplot+[ybar] plot coordinates {(Before,16.84) (After,16.84) };
\addplot+[ybar] plot coordinates {(Before,83.16) (After,8.32) };
\addplot+[ybar] plot coordinates {(Before,0) (After,28.12) };

\legend{\strut Computation, \strut Transfer, \strut Decompression}
\end{axis}
\end{tikzpicture}
\caption{Execution time breakdown before and after applying decompression. Total execution time has reduced by around 47\% after applying decompression}
\label{fig:decompression_overhead}
\end{figure}

Table~\ref{tab:decompressor_performance} shows the decompression performance under P100 using the Brain dataset from \cite{Burtscher:2009:TC}.
\begin{table}[t!]
 \centering
 \caption{Decompression performance on NVIDIA P100 for fixed accuracy and precision modes using the Brain dataset from \cite{Burtscher:2009:TC}}
 \begin{tabular}{|c|c|c|}
  \toprule
  \textbf{Mode} & \textbf{Compression Ratio} $R$ & \textbf{Throughput $W_c$ (GB/s)} \\
  \midrule
  Fixed accuracy & 11.1 & 33.2 \\
  Fixed precision & 9.3 & 35.7 \\
  \bottomrule
 \end{tabular}
 \label{tab:decompressor_performance}
\end{table}
The decoder is able to achieve around 33-35 GB/s with compression ratio $R \approx 10$.
The impact of applying decompression is shown in Figure~\ref{fig:decompression_overhead}.
Applying decompression proves to be beneficial as it reduces the total execution time by 47\%.
However, the decompressor contribution in the total execution time is quite high (52\%).
This high overhead is due to the following reasons:
\begin{enumerate}
 \item The decoding phase in zfp is inherently sequential. The only parallelization that can be achieved is through dispatching "chunks" of compressed bitstream to each CUDA thread. Future work should focus on either: (i) implementing such parallelization and finding the suitable chunk size for the application and underlying HW, or (ii) investigating other "GPU-friendly" compression schemes.
 \item The current CUDA implementation is still experimental and has room for optimizations which were not implemented in this work.
\end{enumerate}
We also believe that the advent of faster and more powerful GPUs will increase the decoder throughput and hence reduce its overhead.

\section{Conclusions} \label{sec:conclusion}
A GPU-based firm real-time system for executing the WHFF model is proposed.
The system is based on COTS HW components.
To run the model within the given constraints, compression, mixed-precision arithmetic, and HBM-enabled GPUs are used.
The system is capable of achieving around 198 GFLOP/s for mixed-precision GEMV.
Data compression enables alleviating the memory bandwidth bottleneck of GPUs. 
Future work should focus on devising GPU-friendly data decompression schemes.

\bibliographystyle{plainurl}
\bibliography{references}

\end{document}